\documentclass[epj]{svjour}
\usepackage{graphics}
\usepackage{color}
\usepackage{amsmath}
\usepackage{graphicx}
\usepackage{braket}
\usepackage{cancel}
\usepackage{float}
\usepackage[table,xcdraw]{xcolor}
\usepackage{multirow}
\usepackage{subcaption}
\usepackage[greek,english]{babel}

\usepackage[square,numbers]{natbib}
\begin{document}

\title{Statistics and systematics of electron EDM searches with BaF}

\author{The NL-$e$EDM collaboration: A. Boeschoten\inst{1,2}, V.R.~Marshall\inst{1,2}, T.B.~Meijknecht\inst{1,2}, A.P.~Touwen\inst{1,2}, P.~Aggarwal\inst{1,2}, N.~Balasubramanian\inst{1,2}, R.~Bause\inst{1,2}, H.~L.~Bethlem\inst{1,3}, A.~Borschevsky\inst{1,2}, T.H.~Fikkers\inst{1,2}, P.A.B.~Haase\inst{1,2}, Y.~Hao\inst{1}, S.~Hoekstra\inst{1,2}, J.W.F.~van~Hofslot\inst{1,2}, S.A.~Jones\inst{1,2}, K.~Jungmann\inst{1}, J.E.J.~Levenga\inst{1,2}, M.C.~Mooij\inst{2,3}, H.~Mulder\inst{1,2}, B.A.~Nijman\inst{1,2}, E.H.~Prinsen\inst{1,2}, B.J.~Schellenberg\inst{1,2}, I.E.~Thompson\inst{1,2}, R.G.E.~Timmermans\inst{1,2}, L.~van~Sloten\inst{1,2}, W.~Ubachs\inst{3}, J.~de~Vries\inst{2,4}, L.~Willmann\inst{1,2}, Y.~Yin\inst{1,2}
}
\institute{Van Swinderen Institute for Particle Physics and Gravity (VSI), University of Groningen, Nijenborgh 3, 9747 AG Groningen, The Netherlands \and Nikhef, National Institute for Subatomic Physics, Science Park 105, 1098 XG Amsterdam, The Netherlands \and Department of Physics and Astronomy, LaserLaB, Vrije Universiteit, De Boelelaan 1100, 1081 HZ Amsterdam, The Netherlands \and Institute for Theoretical Physics, University of Amsterdam, Science Park 904, 1098 XH Amsterdam, The Netherlands }
\date{\today / Received: date / Revised version: date}

\abstract{
The NL-$e$EDM experiment searches for a non-zero electric dipole moment of the electron $d_e$ ($e$EDM) in the ground state of barium monofluoride (BaF). A beam of BaF from a supersonic expansion source is probed with the spin precession method presented in \cite{Boeschoten2024}. This method permits the extraction of an $e$EDM value as well as values for parameters causing a possible systematic bias leading to a false $e$EDM.    
The currently achievable sensitivity is limited by statistics collected in a period of 34 hours and yields an $d_e$ of $2(3) \times 10^{-25}$ $e\,$cm. Furthermore, from the same dataset sufficiently strong limits on parameters which can induce a false $e$EDM are extracted. These are mainly the electric field \textbf{E} and the intensity of the lasers fields in the fiducial volume of the experiment. We summarize the steps required to upgrade of the experiment to reach a competitive level on $d_e$, e.g. an intense laser-cooled beam from a cryogenic buffer gas source and the light collection efficiency of fluorescence. 
%
} 

\authorrunning{NL-$e$EDM \textit{}}
\titlerunning{Statistics and Systematics of $e$ED Searches}
\maketitle
\section{Introduction}
\label{intro}
Searches for permanent electric dipole moments (EDMs) with composite systems, such as atoms and molecules, benefit from their high sensitivity to new interactions between elementary particles that violate both time-reversal (T) and parity (P) symmetry. Since for local quantum field theories T violation is equivalent to the violation of CP -- the combination of charge conjugation (C) and parity -- a nonzero EDM implies P,T and equivalently CP violation \cite{Pospelov2005,Jungmann2013,Engel2013}.

The Standard Model (SM) of particle physics (without neutrino masses) contains two terms that violate CP. It has long been experimentally established that CP is not conserved in flavor-changing weak interactions. In the SM this is parametrized by the complex phase in the Cabibbo-Kobayashi-Maskawa (CKM) quark-mixing matrix. The EDMs that result from this CKM phase are far too small to be detected at present \cite{Khriplovich:1981ca,Hoogeveen:1990cb,Ema:2022yra}. However, the SM also contains the possibility of another, as yet unobserved source of T (CP) violation, which is the term parametrized by the ``QCD vacuum angle'' $\overline{\theta}$ \cite{tHooft:1976snw}. The value of $\overline{\theta}$ is constrained to be less than about $10^{-10}$ by the experimental upper limits on EDMs of the neutron \cite{Abel2020} and the $^{199}$Hg atom \cite{Graner2016}. It is a persistent puzzle why the $\overline{\theta}$ parameter is so small.

New sources of CP violation are expected in scenarios that embed the SM in a more complete theory of particle physics. Such theories generically predict EDM values that are measurable in the ongoing or upcoming EDM experiments. In addition, in order to generate a matter-antimatter asymmetry observed in the universe, CP violation beyond the SM is required \cite{Sakharov1967} assuming CPT to be a good symmetry \cite{Bertolami:1996cq}. For example, in electroweak baryogenesis \cite{vandeVis:2025efm}, this typically leads to EDMs larger than SM predictions. The experimental discovery of EDMs would therefore be a major contribution to the road map towards improved models for particle physics.

 The strongest constraints on the electron EDM ($e$EDM) in the last decade were obtained in paramagnetic diatomic molecules. Experiments with ThO molecules by the ACME collaboration \cite{Andreev2018} and with HfF$^+$ molecular ions by the Boulder group \cite{Roussy2023} have provided new tight limits on the $e$EDM. Our experimental setup is partly similar to the experiment on YbF at Imperial College, London \cite{Hudson2011}. Our approach employs barium monofluoride (BaF) and differs in particular in methods addressing systematic biases in $e$EDM experiments. Experiments with paramagnetic molecules have the potential to improve their sensitivity to the $e$EDM in the coming years by several orders of magnitude. It has been shown recently that paramagnetic molecules provide not only the best limits on the $e$EDM and new CP-violating quark-electron interactions, but they can also be competitive with the neutron and diamagnetic atoms EDM searches in constraining hadronic sources of CP violation \cite{Flambaum:2019ejc, Mulder:2025esr}.
 
This article reports on the implementation of the proposal for the NL-$e$EDM experiment \cite{Aggarwal2018} to search for the $e$EDM with BaF molecules. We present results of initial measurements performed with a beam from a supersonic source of BaF molecules as a proof-of-principle of the spin-precession method \cite{Boeschoten2024}. The method allows for the extraction of a number of systematic biases as well as a robust limit on a possible $e$EDM. We present the current sensitivity, which is limited by statistics, and we discuss ongoing upgrades to reach competitive levels with laser-cooled BaF molecules. We first discuss the setup of the NL-$e$EDM experiment, after which we present the results of the new spin-precession analysis method which was introduced in Ref. \cite{Boeschoten2024}. We conclude with a discussion of the prospects for the next phase of the experiment.

\begin{table}[tb]
\caption{Searches for the $e$EDM in a number of different systems. Listed are the typical coherence times $T$ and the resulting sensitivity in terms of the precession phase $\delta\phi$ \cite{Boeschoten2023}.  
\label{tab:Expsensitivities}}
\centering
\begin{tabular}{l|l|l|l}
\multicolumn{1}{c|}{System} & \multicolumn{1}{c|}{$\delta\phi$} & \multicolumn{1}{c|}{$T$}   & \multicolumn{1}{c}{EDM limit} \\
\multicolumn{1}{c|}{} &\multicolumn{1}{c|}{[rad]} & \multicolumn{1}{c|}{[s]}  & \multicolumn{1}{c}{[$e$~cm] (95\% c.l.)} \\ 
\hline\hline
\multicolumn{4}{l}{\textbf{Beam}}  \\ \hline
 Tl \cite{Regan2002} &    $1\times10^{-6}  $  & $2.4\times10^{-3}$  &  $|d_{\rm{Tl}}|<1.1\times10^{-24} $  \\
ThO \cite{Andreev2018} & $3\times 10^{-6}$ & $1.0\times 10^{-3} $  & $|d_{e}|<1.31\times10^{-29} $   \\
YbF \cite{Hudson2011} & $4\times 10^{-5}$ & $0.7\times 10^{-3}$   & $|d_{e}|<1.27\times10^{-27} $\    \\
BaF   & $1 \times 10^{-3}$ &   $0.8 \times10^{-3}$   &  $|d_{e}|<7 \times 10^{-25}$ \\   \hline
\multicolumn{4}{l}{\textbf{Trap}}  \\ \hline
Ra \cite{Parker2015} & $3\times 10^{-3}$  & 0.035   & $|d_{\rm{Ra}}|<5\times10^{-22} $  \\
Yb \cite{Zheng2022} & $1\times10^{-3}$ & 300 & $|d_{\rm{Yb}}|<1.5\times10^{-26} $    \\
HfF$^+$ \cite{Roussy2023} & $1\times 10^{-3}$& 3  & $|d_{e}|<5\times10^{-30} $ \\ \hline
\end{tabular}
\end{table}

\section{The NL-$e$EDM experiment}
\label{setup}

 \begin{figure}[t]
 \centering
 \resizebox{.5\textwidth}{!}{%
   \includegraphics{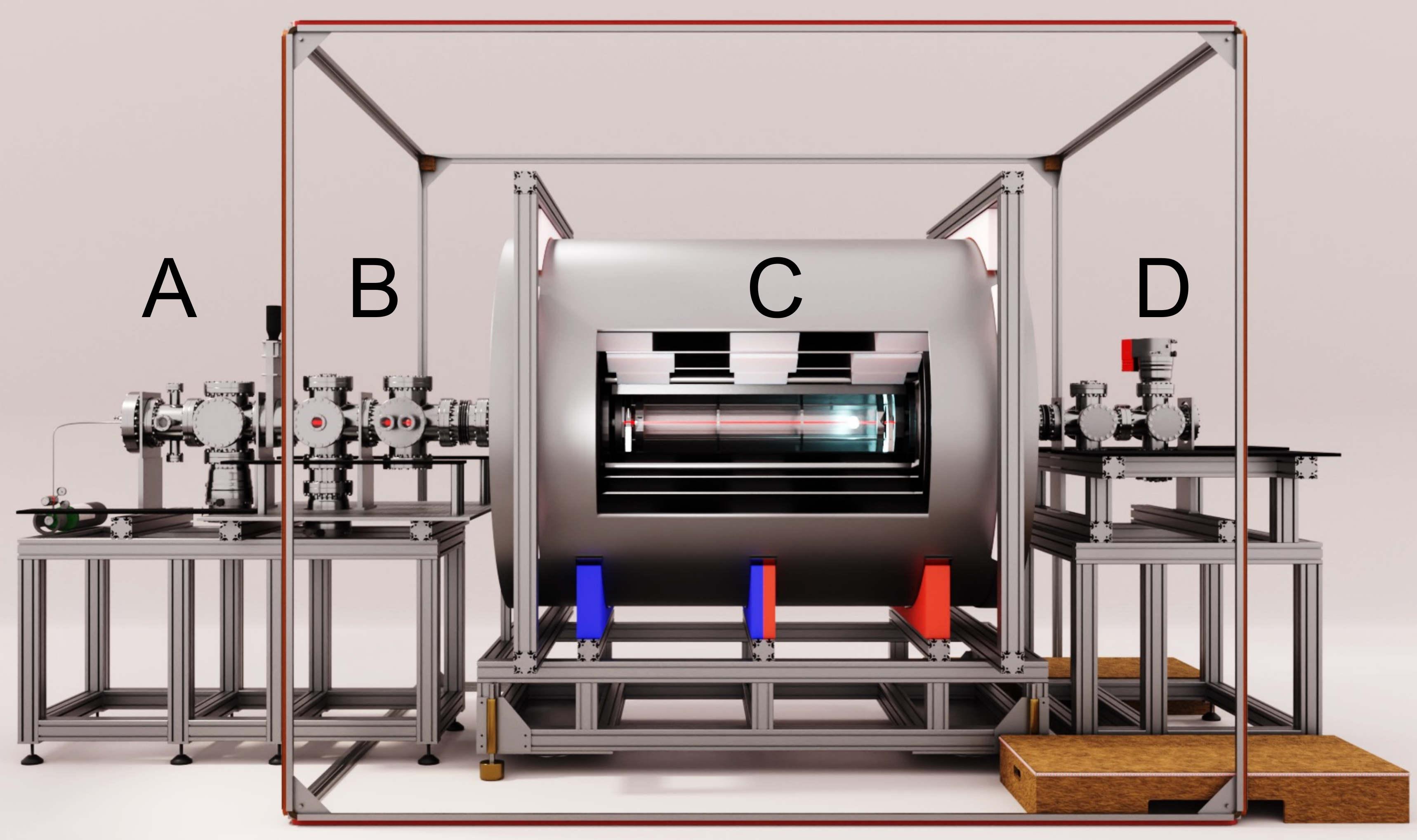}
           }
 \caption{Overview of the modular experimental setup to scale. (A) Supersonic expansion source creating BaF molecules with a velocity of about 600m/s and a narrow velocity spread. The source is operated at repetition rates from 5--30 Hz. (B) Optical probing region for determination of the number of molecules and preparation of the molecules in a specific hyperfine state of the $X ^2\Sigma^+$ ground state. These molecules enter (C) the region in which the spin-precession measurement is performed, consisting of a five layer magnetically shielded volume with magnetic fields of order nT and electric fields of several kV/cm. The environmental fields of order 50~$\mu$T are reduced to less than 5~$\mu$T with rectangular field coils around the region. (D) The spin precession is read out by probing the population distribution in the two hyperfine states of the ground state. 
 }
 \label{fig:overview}
 \end{figure}

The NL-$e$EDM experiment utilizes a molecular beam of BaF and a novel spin-precession method, which provides access to key experimental parameters \cite{Boeschoten2024}. In paramagnetic diatomic molecular EDM  experiments the measurements are sensitive to the induced molecular EDM $D^{P,T}$ \cite{Sandars1965, Sandars1967} which can be written as an enhanced electron EDM, i.e.
$d_eP(E_{\rm{ext}}) W_d  \hbar/E_{\rm{ext}}$, with the molecular CP-violation parameter $W_d$, the molecular polarization $P(E_{\rm{ext}})$ in the external electric field $E_{\rm{ext}}$.  The molecular parameter $W_d$ can be determined by molecular structure calculation.
The size of the enhancement depends strongly on the molecular system and was recently calculated for BaF  \cite{Haase2021}. In some atoms, this enhancement can reach values of $10^3$--$10^4$, while in molecules it can reach up to $10^9$~\cite{Dzuba2012}. 

The statistical sensitivity $\delta d_e$ for the BaF $e$EDM experiment in a molecular beam is determined by the interplay of five quantities: The rate of detected molecules $\dot n = dn/dt$, the spin-precession time $T$, the measurement time $T_{\rm{tot}}$, the degree of polarization of the molecule in an external electric field $P(E_{\mathrm{ext}})$ and the value of the molecular parameter $W_d$ for BaF, which has been determined to 
$W_d=3.13 \pm 0.12\times10^{24}\mathrm{Hz}/e\  \rm{cm}$
\cite{Haase2021}. The sensitivity is given by
\begin{equation}
\label{statistics}
\delta d_e = \frac{1}{ ~W_{d}P({ E}_{\mathrm{ext}})T \sqrt{\dot{n}_PT_{\rm{tot}} }}.
\end{equation}
This uncertainty is derived from a phase determination in a spin-precession measurement $\delta\phi \approx 1/\sqrt{n_P}$ with $n_P$ the total number of phase estimations \cite{Regan2002}. In any experimental realization, this must be matched by an accurate determination of the parameters governing the measurement process \cite{Boeschoten2024}.

We now discuss the spin-precession method with the connection to statistical sensitivity as well as possible systematic bias on an extraction of an $e$EDM limit. Then we show the experimental realization, which enables the exploitation of the method (Fig. \ref{fig:overview}) and the extraction of quantitative results (Sec. \ref{sec:analysis}).

\subsection{Spin-precession method}
\label{setupSpinPrecession}

\begin{figure*}[t]
  \centering
\includegraphics[width=.9\textwidth]{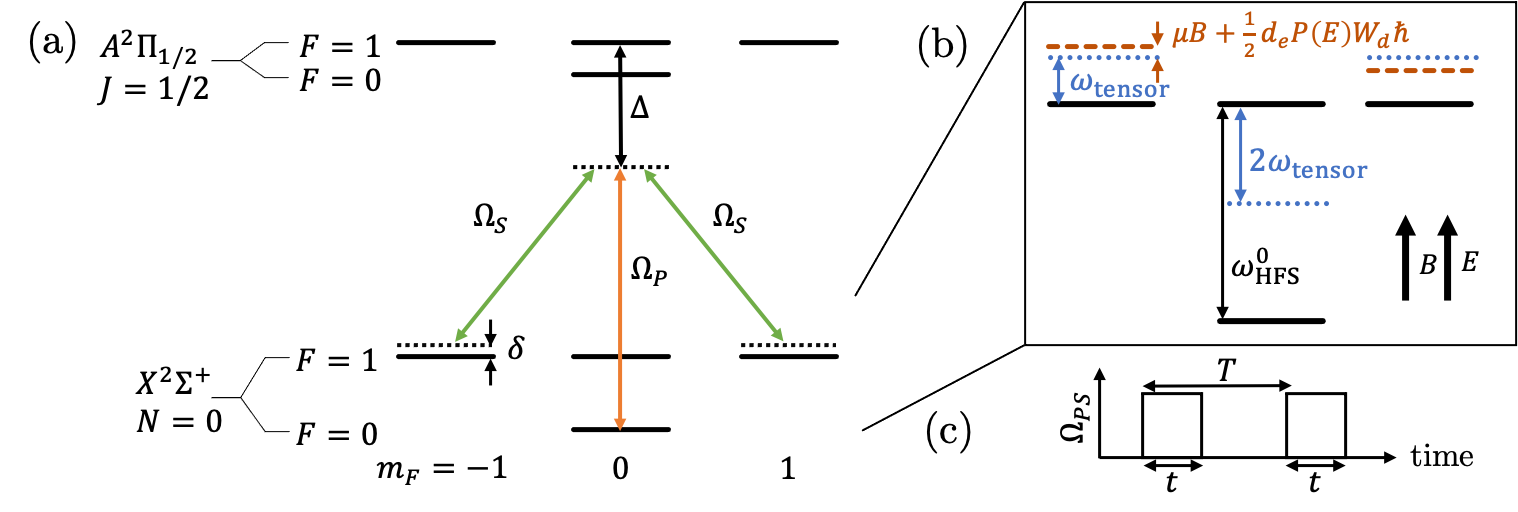}
\caption{The superposition in the $X^2\Sigma^+,v=0,N=0,F=1$ state is created by a two-photon transition via the electronically excited state $A^2\Pi_{1/2},v=0,J=1/2$. The coupling is achieved by two laser fields with Rabi frequencies $\Omega_S$ and $\Omega_P$ at a typical detuning $\Delta=1$~GHz from the $X^2\Sigma^+ - A^2\Pi_{1/2}$ resonance. The detuning $\delta=\omega_{PS}-\omega_{\rm{HFS}}(E)$ is several kHz from two-photon resonance, where $\omega_{\rm{HFS}}(E)=\omega_{\rm{HFS}}^0+\omega_{\text{tensor}}(E)$ and $\omega_{PS}=\omega_P-\omega_S$. (b) The $X^2\Sigma^+,v=0,N=0$ sublevels of the ground state in electric and magnetic fields. The hyperfine splitting in absence of external fields $\omega_{\text{HFS}}^0$ is around $65.85~\text{MHz}$. The tensor Stark shift of the $m_F=\pm1$ levels, $\omega_{\text{tensor}}(E)$, is around $15~\text{kHz}$ at an electric field of 2 kV/cm. The tensor Stark shift of the $m_F=0$ level is approximately twice that of $\omega_{\text{tensor}}$ with opposite sign. 
(c) The timing sequence of the laser-light pulses with effective Rabi frequency $\Omega_{PS}$, where typical pulse lengths are $t = 80~\mu$s and the pulse separation period is $T=1$~ms. Energy levels and timings are not to scale.
}
\label{fig:SpinPrecessionMethod}
\end{figure*}

The measurement is performed in the $X^2\Sigma,v=0,N=0$ ground state of $^{138}$Ba$^{19}$F. The relevant states are hyperfine states $\ket{F,m_F}$ with the angular momentum $F = 0, 1$ and magnetic quantum number $m_F$. 
A superposition of these states is created by a two-photon process (Fig. \ref{fig:SpinPrecessionMethod}) and can be written as  
\begin{equation}
\ket{\psi} = \alpha \, \ket{1,-1} +\alpha'\,\ket{1,1}+\beta\,\ket{1,0}+\gamma \, \ket{0,0}.  
\end{equation}
The coefficients $\alpha, \alpha',\beta$ and $ \gamma$ are determined by the parameters of the two-photon process which are experimentally controlled. The parameters are the two-photon detuning from the hyperfine structure $\delta = (\omega_P-\omega_S)-\omega_{\rm{HFS}}(E)$ in the range of kHz and detuning $\Delta$ from an excited state of order GHz. Furthermore, the length of the two-photon pulse $t$ and the intensities of both laser fields determine the Rabi-frequencies $\Omega_{P/S}$ together with the polarizations of the laser fields $\hat{e}_{P/S}$. These parameters, together with the electric and magnetic fields $\textbf{E}$ and  $\textbf{B}$ determine the evolution of the state $\ket{\psi}$. A second pulse after a time $T$ is applied to project the superposition back to populations in $F=0$ or $F=1$. The probability $P_{0,1}$ of finding the molecule in $F=0$ or $F=1$ state is a function of the parameters,
  \begin{equation}
  \label{eq:SpinPression}
  P_{0,1}=P_{0,1}(\delta, \Delta, t, T, \Omega_{P/S}, \hat{e}_{P/S}, \textbf{E}, \textbf{B}), 
  \end{equation} 
which is numerically determined by solving the set of Optical Bloch equations \cite{BoeschotenThesis} for the relevant eight states in external electric and magnetic fields (Fig. \ref{fig:SpinPrecessionMethod}). The description provides the access to extract experimental parameters from the observed spin-precession signal and reduces the number of additional measurements for the determination of, e.g., the electric field strength, the spin-precession contrast or the Rabi frequencies $\Omega_{P/S}$ as we show Section \ref{sec:analysis}.

\begin{figure}[bt]
  \includegraphics[width=1\linewidth]{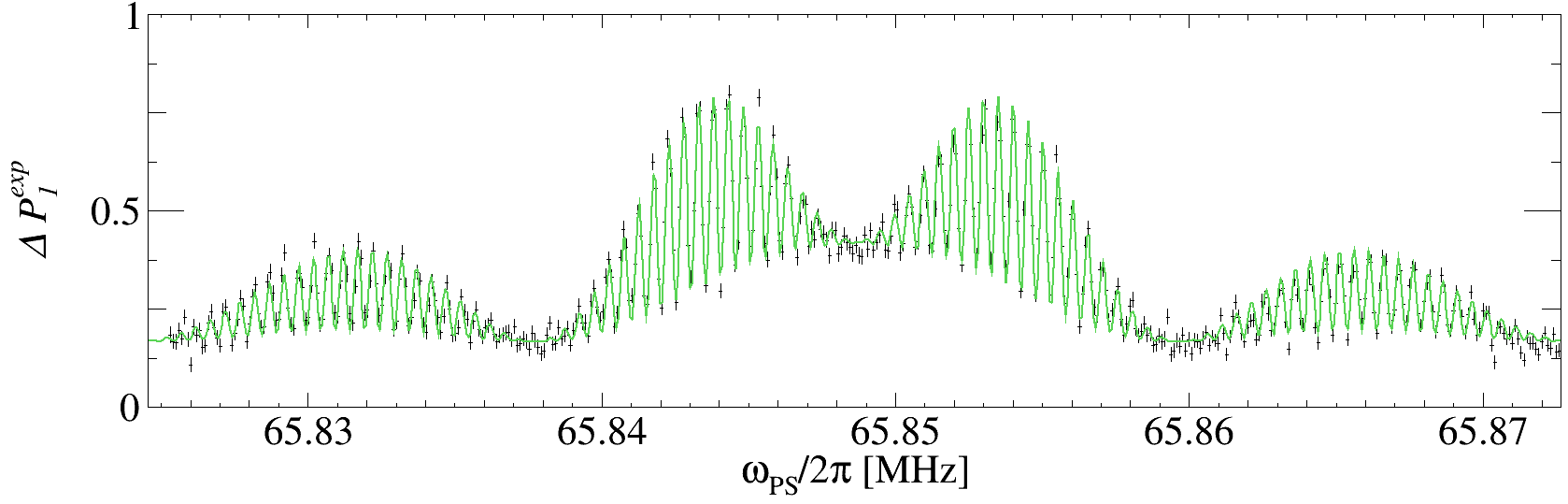}
  \caption{An example of the spin-precession signal $P_0(\delta)$ while keeping the parameters constant ($B = 4.40$~nT, $E  = 0$~kV/cm, $t=80 ~\mu$s and $T= 2$ ms). The green line is according to the model of Eq. \ref{eq:SpinPression}. The hyperfine structure splitting in the ground state is $\omega^0_{\rm{HFS}} = 65.84854(3)$~MHz. The central fringes of the spectrum are most sensitive to a possible $e$EDM.}
  \label{fig:spinpres}
\end{figure}
  
  In electric and magnetic fields the magnetic hyperfine states $\ket{F,m_F}=\ket{1,\pm1}$ acquire a relative phase difference 
  \begin{equation}
  \label{eq:phase}
      \phi =  (2\mu B/\hbar \pm d_e P(E_{\mathrm{ext}}) W_{d}) T,
  \end{equation} 
  where the sign in front of the $e$EDM contribution depends on the relative orientation between the electric and magnetic field vector.  The limit on the EDM of the BaF molecule is extracted from pairs of spin-precession measurements where the relative directions of the two fields have been reversed.
  Since the phase difference is extracted from measurements of the signal $P_{0,1}$ (Eq. \ref{eq:SpinPression}), the dependence of $P_{0,1}$ on the experimental parameters has to be taken into account. We show that the experimental procedure permits the measurement of these parameters with sufficient precision to constrain a systematic bias on the $e$EDM determination.

\subsection{Experimental implementation}
\label{sec:Implementation}

\subsubsection{Molecular beam source}
Barium monofluoride (BaF) molecules are produced in a supersonic source. The source uses a rotating 3~mm wide Ba disk with 40~mm diameter. Ba atoms are ablated by a pulsed Nd:YAG laser and seed a carrier gas of $2 \%~\textrm{SF}_{6}$ and $98\%$~argon from a pulsed Even-Lavie valve. The source has a variable repetition rate typically set to 10~Hz and produces a molecular beam with a rotational temperature of 3.5~K. 
The mean velocity of the molecules is $600$~m/s, with a velocity spread of about $35$~m/s \cite{Aggarwal2021}. The average intensity is $3.3\times10^9$ molecules/sr/pulse, which is a factor $\sim 5$ higher compared to the intensity reported in~\cite{Aggarwal2021}.
The molecular beam pulse passes through a skimmer (diameter 5~mm) 28~cm after the exit of the Even-Lavie valve. Differential pumping provides for a vacuum pressure below $10^{-7}$~mbar downstream. 

\subsubsection{Control of magnetic and electric fields} \label{beamsource}
\begin{figure}
\centering
\resizebox{0.4\textwidth}{!}{%
  \includegraphics{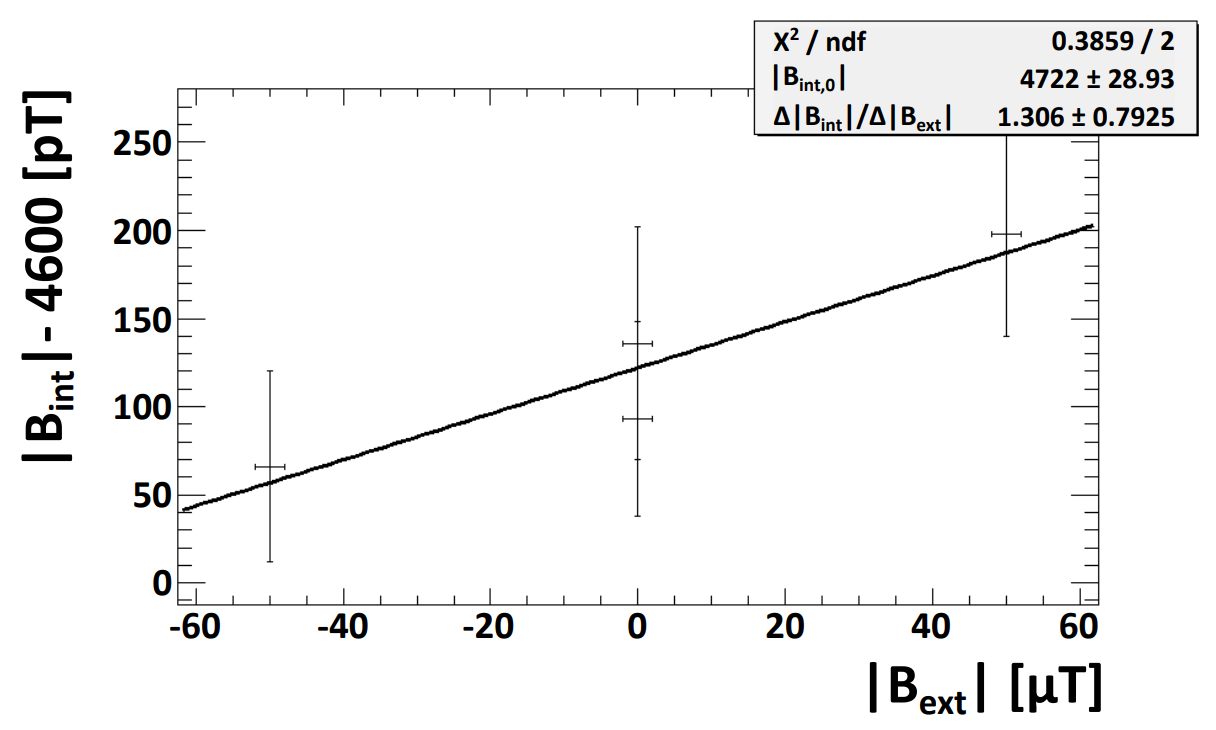}
}
\caption{The magnetic field inside of the shielding is derived from spin-precession signals when the magnetic shield was exposed to a magnetic field change of $\pm50 ~\mu\mathrm{T}$. The field change inside the shielding was suppressed by $1.3(8) \times 10^{-6}$ in agreement with numerical simulations in COMSOL.}
\label{fig:Shielding}
\end{figure}

A multilayer magnetic shielding has been constructed in order to provide a stable near-zero-field volume for performing the spin-precession measurement.
The mu-metal shielding is designed with numerical simulation methods (COMSOL) in order to suppress external fields by 6 orders of magnitude. The fiducial volume has a 0.5~m diameter and 1.2~m length \cite{MeijknechtThesis}. Additional suppression of external fields is achieved by large coils surrounding the shield, which compensate environmental fields in the laboratory of about 50~$\mu$T by a factor of 10. The magnetic field strength outside of the magnetic shield is constantly monitored with a resolution of better than $10$~nT/$\sqrt{\rm Hz}$. Variations of the external magnetic field are less than 100~nT over the course of the day. The suppression of external magnetic fields has been determined by exposing the shield to a bias field of $\pm ~ 50~\mu$T externally while measuring the magnetic field field change inside of the shielding with the spin-precession signal from the BaF molecules (Fig. \ref{fig:Shielding}). This suppression of six orders of magnitude leads to a magnetic field change of less than $100$~fT in the course of a day.   

The design of structures for generation of homogeneous $\textbf{E}$ and $\textbf{B}$ fields has been assisted by simulation (COMSOL) \cite{MeijknechtThesis}. The magnetic holding field orthogonal with respect to the molecular beam of order 10~nT is generated by a cylindrical structure of 0.3~m diameter and length 1~m. A current distribution of $I(\theta) = I_0 \cos(\theta)$, with $\theta$ the azimuthal angle around the cylinder is generated by adjusting the wire density along the cylinder. This is known as a cosine-coil. The electric field parallel to the magnetic fields is generated by glass plates with a conductive coating (ITO) of 750~mm length and 100~mm height. The field plates are separated by 40~mm in a rigid construction of machinable ceramics (Macor) and glass. 

The high voltage is provided by a positive and a negative high voltage supply (ISEG HPp/n 300 106). The electric field direction is reversed by a set of reed relays controlled via TTL signals from the experiments data and control system (DAQ) system. This permits a variable sequence of electric field direction and zero field while the power supplies are kept at constant voltage throughout the measurements in order to avoid time-dependent electric fields due to the limited voltage ramping rates. 
The field strength of several kV/cm is measured to an accuracy of order 1~V/cm with the spin-precession signal $P_{0}(\delta)$. This provides an absolute measurement of the field strength in the fiducial volume \cite{Boeschoten2024}. 

\subsubsection{Laser frequency and stability}
\label{sec:lasersystem}

\begin{figure}
    \centering
    \includegraphics[width=0.5\textwidth]{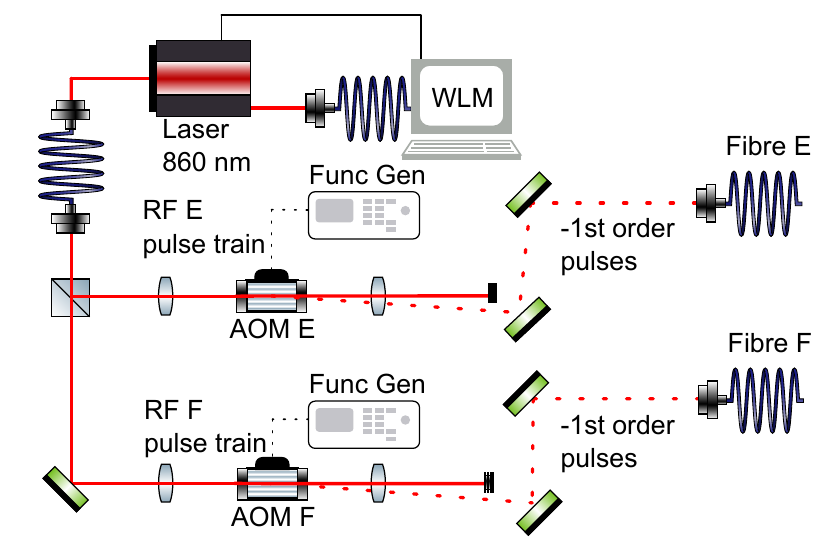}
    \caption{Pulses with precise timing and phase coherence for the spin-precession methods are generated from light of one laser at 860~nm (TOPTICA TApro). The light is split by a beam splitter into two beams. Each beam is passed through an AOM which is driven by RF pulses generated DDS function generators referenced to a GPS controlled Rb-clock. The pulsed 1st order beams of each AOM are aligned into optical fibers E and F.}
    \label{fig:Pulse_generation}
\end{figure}

The light beams at multiple optical frequencies required in the experiment are provided by diode lasers and Ti:Sapph systems. The frequencies are controlled by a HighFinesse WS8-2 wavelength meter \cite{WLMWS8-2} in combination with a HighFinesse 8 channel photonic crystal switch to control up to 8 wavelengths simultaneously. A frequency stability sufficient for an experiment of better than 0.5~MHz over long periods ($>$ 10h) is achieved \cite{MarshallThesis}. 
The laser sources provide continuous wave output. Acousto-optical modulators (AOMs) are employed to create pulses with controlled frequency offset and sub-ns timing. The RF-frequencies and the timing are derived from direct digital synthesis function generators, which are referenced to a GPS stabilized Rb-clock (SR FS725).
Laser light at 815~nm is used for the determination of the molecular beam intensity on the $X^{2} \Sigma^{+} (v=0,N=1)$ -- $A^2\Pi_{3/2}(v=0,J=3/2)$ transition by laser induced fluorescence in Section B (Fig. \ref{fig:overview}) and counts per molecular pulse $n_{\rm{norm}}$ is recorded by the DAQ. Rotational pumping from the $X^{2} \Sigma^{+} (v=0)$ $N=2$ state to $N=0$ state and hyperfine optical pumping of the $X^{2} \Sigma^{+} (N=0)$ $F=1$ to $F=0$ state employs light at 860~nm also in Section B.
The population in the $X^{2} \Sigma^{+} (v=0,N=0, F=1)$ state after the spin precession is probed with the $X^{2} \Sigma^{+},N=0, F=1$ -- $A^2\Pi_{3/2},J=3/2, F=1$ transition in Section D (Fig. \ref{fig:overview}). The counts $n_{\rm{sp}}$ is also recorded to allow for the experimental determination of $P_1$ (Eq. \ref{eq:SpinPression}).

The number of detected photons is ${n_{\rm sp}} = \epsilon \ n_{\rm{molecules}}$, where the detection efficiency $\epsilon = 7 \times 10^{-4}$ results from the solid angle of light collection and the quantum efficiency of the detector. The counts are recorded within timing bins of $100~\mu\rm{s}$, which corresponds to 10~m/s resolution on the velocity of the molecules. Integrated over the velocity profile of the beam, we observe about 100 photons per molecular pulse. However, the average yield during the data taking was 20 photons/pulse due to the degradation of the molecular yield from the source over the course of several hours. 

The laser pulses for the superposition creation and readout (see Fig. \ref{fig:SpinPrecessionMethod}) are derived from a single laser at a wavelength of 860nm with a pair of acousto optical modulators (AOMs) (Fig. \ref{fig:Pulse_generation}). These light fields for the coupling to the spin-precession state are delivered by beams counter-propagating to the molecular beam, providing for light pulse intensity (Rabi frequency $\Omega_{PS}$) and timing ($t$ and $T$) independent of the molecular velocity.  Parameters such as laser frequency, intensity, pulse lengths and timing are recorded by the data acquisition system with the spin-precession data for every single molecular pulse. 

\subsubsection{From fluorescence detection to spin-precession signal}

The signal counts per molecular pulse $n_{\rm{norm}}$ provides a measurement of the number of produced molecules, while the counts $n_{\rm{sp}}$ detected during the time of flight window between 6 and 7~ms yields the number of molecules in the hyperfine state $F=1$ after spin precession.  
The spin-precession $P_1$ (Eq. \ref{eq:SpinPression}) is determined in the experiment by
\begin{equation}
    P_1^{\rm{exp}} = n_{\rm{sp}}/(f \cdot n_{\rm{norm}}),
\end{equation} 
with a calibration factor $f$ which is determined when we do not drive the spin precession.
The sum of the populations in the hyperfine states $\ket{1,-1}$ and $\ket{1,1}$ is determined with a statistical uncertainty which is given by counting statistics. The uncertainty of a measurement, e.g. Fig. \ref{fig:bshift}, is determined by counting statistics including the background from scattered light. 

\section{Spin-precession analysis}
\label{sec:analysis}

\begin{figure}[bt]
  \includegraphics[width=1\linewidth]{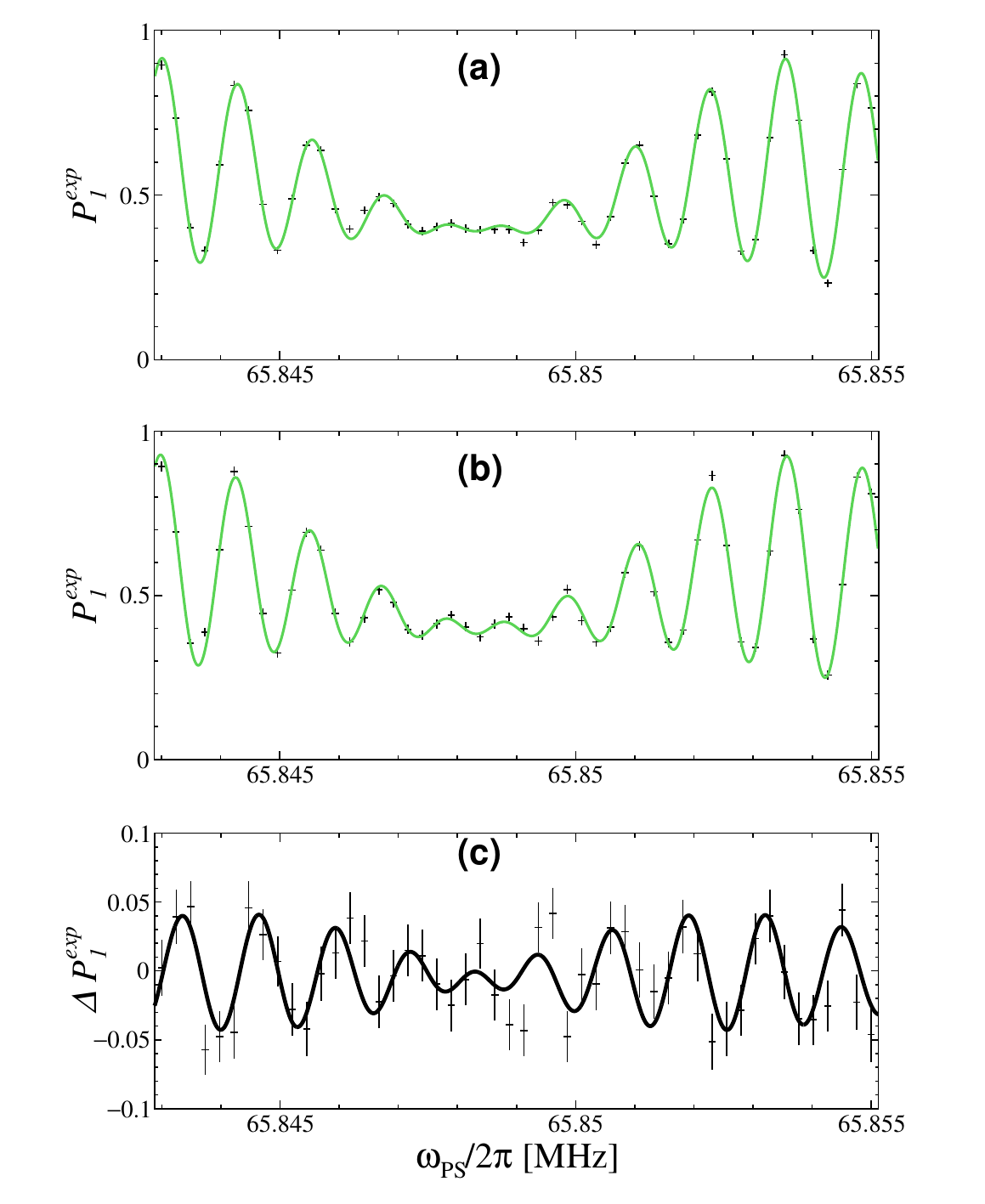}
  \caption{Spin-precession signals for (a) molecules with velocity $600\pm6$~m/s and (b) $576\pm6$~m/s. The Rabi frequency $\Omega_{PS}$ differs due to the Doppler shift contribution to the detuning $\Delta$. (c) The change of $\Delta$ of 24~MHz causes a 3.0(5)\% change in $\Omega_{PS}$ which results in a dependence of $\Delta P_1^{\rm{exp}}$ on the frequency $\omega_{PS}$ .
  }
  \label{fig:rabishift}
\end{figure}
\begin{figure}[bt]
  \centering
  \includegraphics[width=1\linewidth]{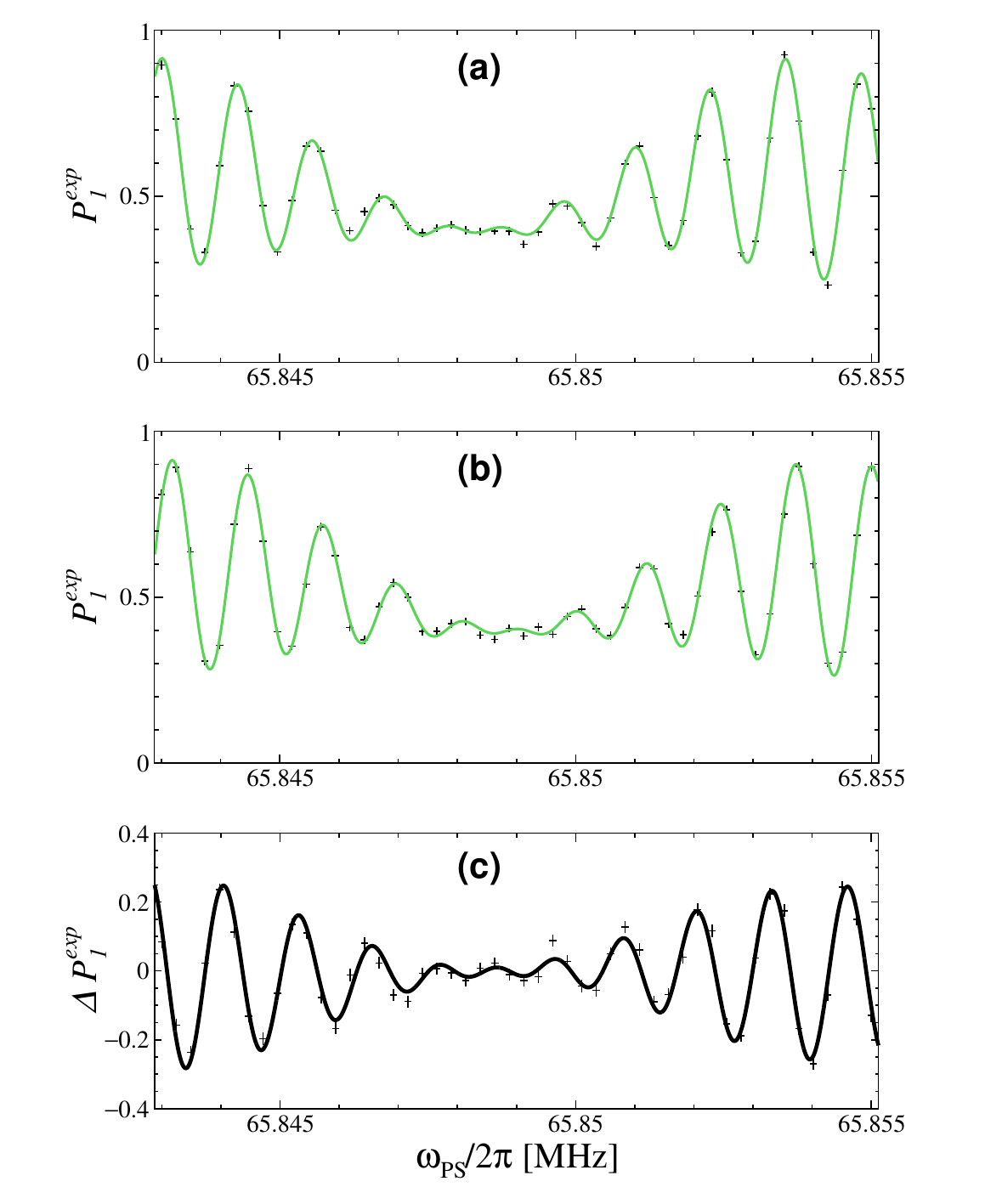}
  \caption{Spin-precession signals for (a) electric field parallel and (b) antiparallel to the magnetic field. The electric field strength was different by about 1 \%. (c) The difference in field strength causes a dependence of $\Delta P_1^{\rm{exp}}$ on the frequency $\omega_{PS}$. A change of field strength of $-23.0(2)$~V/cm at a field of 1.970~kV/cm is extracted.}
  \label{fig:eshift}
\end{figure}
\begin{figure}[bt]
  \centering
  \includegraphics[width=1\linewidth]{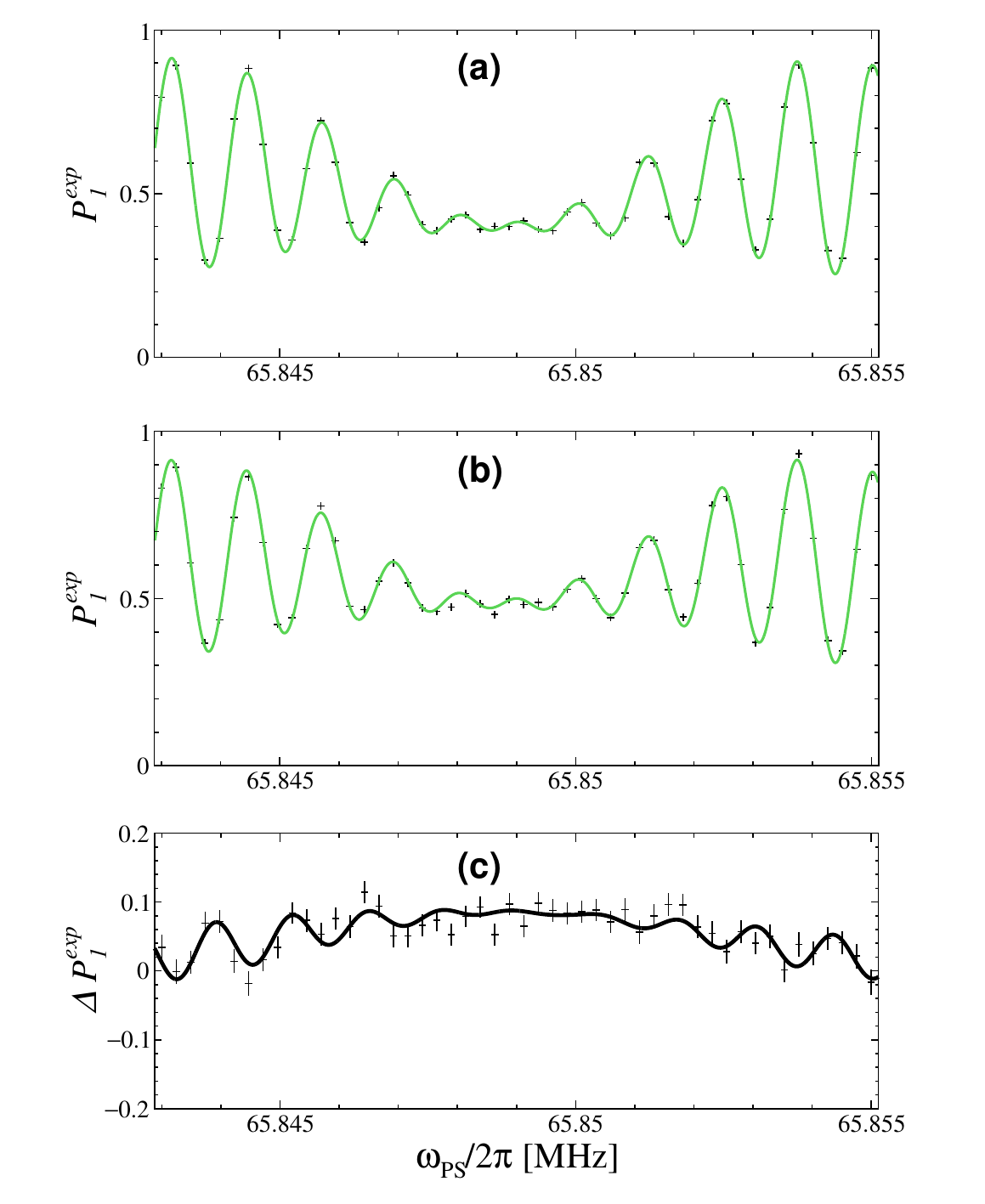}
  \caption{Spin-precession signals for (a) a magnetic holding field of $-4.71(1)$~nT and (b) 5.46(1)~nT. The two spectra were recorded 23~h apart.
  (c) The difference of the signal $\Delta P_1^{\rm{exp}}$ yields a difference in magnetic field strength of 0.72(3)~nT and a change of 1.0(4)\% in $\Omega_{PS}$ for these two datasets.}
  \label{fig:bshift}
\end{figure}

\begin{table}
    \centering
    \begin{tabular}{c|c|l}
    Parameter &  $\#$ Values &  Parameter Range  \\
    \hline
    Timing & 1 & $t = 80~\mu\rm{s} $, $ T= 1~\rm{ms}$ \\
     $\delta$      & 50 & $\pm$ 6~kHz   \\
     $\textbf{E}$ & 2  & $ -1.947(1)\, \rm{and} +1.970(1)~\mathrm{kV/cm}$ \\
     $\textbf{B}$ & 2 & $ -4.715(12)~\mathrm{nT}, +5.462(12)~\mathrm{nT}$   \\
    $\Omega_{PS}$  & 5 & $\pm 3\%$   \\
$n_{\rm{sp}}$ & $10-100$ & per ablation pulse \\
    \end{tabular}
    \caption{List of the different experimental conditions during the collection of data in about 34~h.}
    \label{tab:dataset}
\end{table}

Data collected within 2 days is evaluated for extracting the current sensitivity of the experimental setup. Parameters during the measurement were set at the start and data collection continued automated (Table \ref{tab:dataset}). During the data taking, the values for the detuning 
$\delta$ span over 12 kHz in 50 steps in order to record spin-precession signals $P_1(\delta)$ such as in Figs. \ref{fig:rabishift}, \ref{fig:eshift} and \ref{fig:bshift}. The electric field direction was reversed every 150~s, which corresponds to 50 steps in $\delta$. The magnetic field reversal was done only after 24~h in order to reduce effects from hysteresis in the magnetic shielding. The field reversal resulted in a different field strength. 
Parameters related to the timing of the pulse sequence are set to an accuracy better than 1~ns by function generators to a low phase noise which are referenced to a GPS disciplined Rb-clock. Frequencies of all the lasers are controlled by a wavelength meter to a precision of better than 1~MHz. Parameters of the measurement (Table~\ref{tab:spresults}) are either controlled to the required accuracy or are extracted from the observed spin-precession fitted against the model function $P_1$ (Eq. \ref{eq:SpinPression}).  

The $e$EDM contribution is derived from the difference of measurements with parallel and anti-parallel electric and magnetic fields.
The magnetic field strength is chosen to provide for a phase around $\phi = \pi/2$ (Eq. \ref{eq:phase}) and a phase error of $\delta\phi \approx \delta P_1^{\rm{exp}}$ can be extracted. A measurement at a single detuning $\delta \approx 0~\mathrm{kHz}$ would require independent experimental determination of other parameters such as \textbf{E}, $\Omega_{PS}$ and the contrast of the signal for a reliable extraction of an \textit{e}EDM limit. 
The advantage of the method presented here is that the signal Eq. \ref{eq:SpinPression} provides sensitivity to all these parameters. We generate a set of spectra from one dataset (Figs.~\ref{fig:rabishift}, \ref{fig:eshift} and \ref{fig:bshift}), from molecular pulses which have specific combinations of the parameters such as \textrm{E}, \textrm{B} and $\Omega_{PS}$ at the time of analysis. The spectra are analyzed with the spin-precession model $P_1(\delta)$ with a typical reduced $\chi^2$ in the range of $0.9-1.3$. The precision of the parameters, including the contrast are limited by the same statistics and are sufficient to constrain a systematic bias on the $e$EDM (Table \ref{tab:spresults}). The results presented in this paper are based on an effective measurement time of 17~h for each direction of the magnetic field. The electron $e$EDM $d_e$ = $2(3) \times 10^{-25}\,e$~cm has been extracted from the dataset and the sensitivity is in agreement with the collected statistics.

\begin{table}
    \centering
    \begin{tabular}{c|c|l|l}
    Parameter & Constraint &  $\delta d_e < 10^{-27} e$\,cm & Source\\
    \hline
     $t$      & $<1$~ns  & $ 5$~ns   & clock\\
     $T$      & $<1$~ns  & $ 100$~ns  & clock\\
     $\delta$ & $<1$~Hz  & $ 16$~Hz   & clock\\
     $\Delta$ & $<1$ MHz &  5 MHz  & WLM\\
    \textbf{E}     & $dE/E <2 \times 10^{-3}$ & $<2 \times 10^{-3}$ & stat.\\
    \textbf{B}     & $\delta B <10$~pT & $ < 0.1$~pT  & stat.\\
    $\Omega_{PS}$  & $d\Omega/\Omega < 5 \times 10^{-3}$ & $< 2 \times10^{-3}$  & stat.\\
    \end{tabular}
    \caption{Summary of parameters controlled the experimental implementation (pulse length $t$, spin-precession time $T$ and frequency $\delta$) together with measured parameters, such as laser frequencies controlled by wavelength meter (WLM) and statistical limits of electric and magnetic field and the Rabi frequency. The accuracy of the latter depend on the same statistics as the limit on the $e$EDM. The third column gives the requirements to constrain a systematic bias on a limit on $d_e$. The last column give the method by which the constraint can be achieved.}
    \label{tab:spresults}
\end{table}

\section{Prospects for the next phase}

Various improvements to the experimental configuration can be implemented -- or are already in the process of being implemented -- to increase the statistical sensitivity of an $e$EDM-search represented by Eq.~(\ref{statistics}). In particular, by using slower molecules we increase the spin-precession $T$. In phase 2 of our experiment we replace the supersonic beam source with a laser-cooled, hexapole-focused, cryogenic buffer gas beam~\cite{hofslot2025}. 

In a cryogenic buffer gas beam (CBGB) source~\cite{Hutzler2012}, molecular radicals are produced inside a cold cell via laser ablation of a solid target. The molecules are cooled through collisions with continuously flowing helium or neon buffer gas. Entrained in this gas, they are extracted from the cell to form a cold molecular beam. Based on the design in ~\cite{Truppe2018}, we have constructed a cryogenic buffer gas source that generates pulses containing typically $2\times10^{10}$ BaF molecules in the state $N=0$, with a brightness of $1\times10^{11}$ molecules per sr per pulse at a velocity of 200~m/s~\cite{Mooij2024}.
To create an intense, collimated beam, we will use a combination of an electrostatic hexapole lens, which provides a position-dependent force, and transverse laser cooling, which provides a velocity-dependent friction force. 
Electrostatic multipole lenses have been widely used for state selection and focusing of light polar molecules~\cite{Meerakker2012}. 
We have recently demonstrated~\cite{Touwen2024} that a properly designed hexapole lens significantly enhances the intensity of the downstream BaF beam.

BaF, like its lighter homologous CaF and SrF, possesses a structure amenable to laser cooling. Calculations predict that the ground vibrational level of $A^2\Pi_{1/2}$ decays to the ground vibrational level of $X^2\Sigma^+$ with a branching ratio of 0.964~\cite{Hao2019}. We recently reported the first demonstration of transverse Doppler cooling of a hexapole-focused BaF beam~\cite{hofslot2025}. Using three tunable lasers with appropriate sidebands and detuning, the molecules scattered approximately 400 photons, limited by scattering rate and the available interaction time. The leakage of the population to dark states was less than 10$\%$. The observed scattering rate was 14$\%$ of the theoretical maximum, but this can likely be improved by optimizing sideband power and refining the magnetic field in the cooling region.
The experimental results also served to benchmark trajectory simulations, which we use to predict the achievable flux for the $e$EDM experiment. Our simulations indicate that incorporating the hexapole lens together with the laser cooling stage could increase the molecular flux by two orders of magnitude with respect to the supersonic beam source.
The detection of the signal $n_{\rm sp}$ will be upgraded from the $X^2\Sigma^+ - A^2\Pi_{3/2}$ transition to the $X^2\Sigma^+ - D^2\Sigma^+$ transition at a wavelength of 405~nm. The excited state $D^2\Sigma^+$ primarily decays via the $A^2\Pi$ states by emission of two infrared photons. In addition, the detection will change from a photomultiplier to an avalanche photodiode based detection with an increased solid angle for the light collection.

A comparison between the current `phase 1' experiment and the next stage `phase 2' experiment is given in Table~\ref{tab:Sensitivity1}. The considerable increase in the projected average photon rate (s$^{-1}$) is the consequence of significant combined improvements in the source intensity, the implementation of transverse laser cooling, and an improved fluorescence detection scheme. Because of the increased length of the molecular pulse from the cryogenic source the fraction of molecules taking part in the spin precession will be reduced by a factor 3. Combined, these factors leads to the indicated average photon counting rate. The square root of the photons counts in a day, multiplied with the increased polarization factor in a higher electric field strength, the increased precession time, the interference contrast and the $W_d$ parameter of the BaF molecule (see Equation~\ref{statistics}) leads to the associated statistical sensitivities ($e ~\rm{cm}  ~\mathrm{day}^{-1}$) that are given at the bottom of the table.

\begin{table*}[t]
\caption{Summary of achieved performance of the parts of the experiment (phase 1) and the projected gain in statistical sensitivity for phase 2.}
\label{tab:Sensitivity1}       
\begin{tabular*}{\textwidth}{p{160pt} @{\extracolsep{\fill}} p{60pt}p{100pt}p{200pt}p{120pt}}
\hline\noalign{\smallskip}

Item & Phase 1 & Phase 2 & Comment \\
\noalign{\smallskip}\hline\noalign{\smallskip}
\bf{Molecular Beam Source}\\
Source type& supersonic~\cite{Aggarwal2021} &  cryogenic buffer gas~\cite{Mooij2024} & \\
Brightness (molecules/sr/pulse)& $3.3 \times 10^{9}$ & $1\times 10^{11}$  & in $N=0$\\
Repetition rate (Hz) &10&20&\\
Forward velocity (m/s) & $600\pm30$  & $200\pm 30$ & velocity ($\mathrm{m/s}\pm 1\sigma$ spread) \\
 \hline\noalign{\smallskip}
\bf{Laser Cooling}\\ 
Transverse cooling & 1 & 200 &  gain hexapole and laser cooling \\
Rotational cooling & 1 & 10 &  gain optical pumping\\
 \hline\noalign{\smallskip}
\bf{Spin Precession Parameters}\\
Electric Field (kV/cm)& 2 & 10 &  \\
Polarization factor $P(E_{\rm{ext}})$ & 0.16 & 0.53 & resulting from larger field strength  \\
Precession Time (ms) & $1$ & 3 & Due to reduced velocity\\
Precession fraction &0.9&0.3& Due to length of molecular pulse\\
Interference contrast & 0.9 & 0.9 & \\
\hline\noalign{\smallskip}
\bf{Fluorescence Detection}\\
Signal $n_{\rm{sp}}$ & 1 & 1.5 &  Change detection via $\Pi_{3/2}$ to $D^{2} \Sigma^{+}$ state \\
Light collection & 0.02 & 0.20 & Solid angle of light collection \\
Quantum efficiency& 0.07 & 0.80 & Change from PMT to APD \\
 \hline\noalign{\smallskip}
 \hline\noalign{\smallskip}
Average photon rate (s$^{-1}$) &$200$& $2 \times 10^9$& \\ 
Statistical sensitivity ($e\,$cm day$^{-1}$) &$5\times10^{-25}$& $2\times10^{-29}$  & &\\
\noalign{\smallskip}\hline
\end{tabular*}
\end{table*}

\section{Conclusion}

The spin-precession technique developed in the NL-$e$EDM \cite{Boeschoten2024} demonstrated a statistically limited determination of an $e$EDM  $d_e$ = $2(3)\times 10^{-25}$ $ e$~cm in the first phase. Additionally, the method allows for a determination of experimental parameters such as Rabi frequency $\Omega_{PS}$ and electric field strengths from the same dataset. Their uncertainty permits limiting of the systematic bias to well below the $e$EDM limit. Improved statistics in future measurement will further limit the $e$EDM and the uncertainties in the experimental parameters.

It is attractive to use a longer spin sprecession time $T$, as the sensitivity improves linearly with this parameter. Until recently, increasing the interaction time was accompanied by a significant decrease in the counting rate $\dot N$. However, recent advances in decelerating molecular beams \cite{Osterwalder2010, Bulleid2012, Quintero-perez2014,Vandenberg2014} combined with spectacular progress in molecular laser cooling \cite{Shuman2010} and the demonstration of intense cryogenic molecular beam sources \cite{Patterson2007} have opened a route to circumvent this limitation and make long interaction times possible. Ultimately, we aim perform our experiment on barium-containing molecules in a fountain~\cite{Cheng2016} or in an optical lattice trap~\cite{Bause2025}. A similar approach has recently been suggested for the YbF molecule \cite{Lim2017}.

\section{Acknowledgments}
This work is financed by the Netherlands Organisation for Scientific Research (NWO) as part of the research programme `Physics beyond the Standard Model with cold molecules' with  project number 166, project VI.C.212.016 and project OCENW.XL21.XL21.074.

\section{Author contributions}
All authors have contributed through discussions and corrections to the final manuscript.

%
\bibliography{BaF-bib}

@article{Ema:2022yra,
    author = "Ema, Y. and Gao, T. and Pospelov, M.",
    title = "{Standard Model Prediction for Paramagnetic Electric Dipole Moments}",
    eprint = "2202.10524",
    archivePrefix = "arXiv",
    primaryClass = "hep-ph",
    reportNumber = "UMN-TH-4115/22, FTPI-MINN-22-06",
    doi = "10.1103/PhysRevLett.129.231801",
    journal = "Phys. Rev. Lett.",
    volume = "129",
    number = "23",
    pages = "231801",
    year = "2022"
}

@article{Khriplovich:1981ca,
    author = "Khriplovich, I. B. and Zhitnitsky, A. R.",
    title = "{What Is the Value of the Neutron Electric Dipole Moment in the {Kobayashi-Maskawa} Model?}",
    reportNumber = "IYF-81-115",
    doi = "10.1016/0370-2693(82)91121-2",
    journal = "Phys. Lett. B",
    volume = "109",
    pages = "490--492",
    year = "1982"
}

@article{Bertolami:1996cq,
    author = "Bertolami, O. and Colladay, Don and Kostelecky, V. Alan and Potting, R.",
    title = "{CPT violation and baryogenesis}",
    eprint = "hep-ph/9612437",
    archivePrefix = "arXiv",
    reportNumber = "DF-IST-4-96, IUHET-349",
    doi = "10.1016/S0370-2693(97)00062-2",
    journal = "Phys. Lett. B",
    volume = "395",
    pages = "178--183",
    year = "1997"
}

@article{tHooft:1976snw,
    author = "'t Hooft, G.",
    editor = "Shifman, Mikhail A.",
    title = "{Computation of the Quantum Effects Due to a Four-Dimensional Pseudoparticle}",
    reportNumber = "PRINT-76-0551 (HARVARD)",
    doi = "10.1103/PhysRevD.14.3432",
    journal = "Phys. Rev. D",
    volume = "14",
    pages = "3432--3450",
    year = "1976",
    note = "[Erratum: Phys.Rev.D 18, 2199 (1978)]"
}

@article{Hoogeveen:1990cb,
    author = "Hoogeveen, F.",
    title = "{The Standard Model Prediction for the Electric Dipole Moment of the Electron}",
    reportNumber = "DESY-90-006, ITP-UH-14-89",
    doi = "10.1016/0550-3213(90)90182-D",
    journal = "Nucl. Phys. B",
    volume = "341",
    pages = "322--340",
    year = "1990"
}

@article{vandeVis:2025efm,
    author = "van de Vis, J. and de Vries, J. and Postma, M.",
    title = "{Bubble Trouble: a Review on Electroweak Baryogenesis}",
    eprint = "2508.09989",
    primaryClass = "hep-ph",
    reportNumber = "CERN-TH-2025-161, Nikhef 2025-012",
    year = "2025",
    journal= {ArXiv},
    pages={2506.19069},
    archivePrefix={arXiv},
    primaryClass={physics.atom-ph},
    url={https://arxiv.org/abs/2506.19069},
}

@article{Mulder:2025esr,
    author = "Mulder, H. and Timmermans, R. G. E. and de Vries, J.",
    title = "{Probing the QCD $ \overline{\theta} $ term with paramagnetic molecules}",
    eprint = "2502.06406",
    archivePrefix = "arXiv",
    primaryClass = "hep-ph",
    doi = "10.1007/JHEP07(2025)232",
    journal = "JHEP",
    volume = "07",
    pages = "232",
    year = "2025"
}

@article{Flambaum:2019ejc,
    author = "Flambaum, V. V. and Pospelov, M. and Ritz, A. and Stadnik, Y. V.",
    title = "{Sensitivity of EDM experiments in paramagnetic atoms and molecules to hadronic CP violation}",
    eprint = "1912.13129",
    archivePrefix = "arXiv",
    primaryClass = "hep-ph",
    doi = "10.1103/PhysRevD.102.035001",
    journal = "Phys. Rev. D",
    volume = "102",
    number = "3",
    pages = "035001",
    year = "2020"
}

@article{Abel2020,
  title = {Measurement of the Permanent Electric Dipole Moment of the Neutron},
  author = {Abel, C. and Afach, S. and Ayres, N. J. and Baker, C. A. and Ban, G. and Bison, G. and Bodek, K. and Bondar, V. and Burghoff, M. and Chanel, E. and Chowdhuri, Z. and Chiu, P.-J. and Clement, B. and Crawford, C. B. and Daum, M. and Emmenegger, S. and Ferraris-Bouchez, L. and Fertl, M. and Flaux, P. and Franke, B. and Fratangelo, A. and Geltenbort, P. and Green, K. and Griffith, W. C. and van der Grinten, M. and Gruji\ifmmode \acute{c}\else \'{c}\fi{}, Z. D. and Harris, P. G. and Hayen, L. and Heil, W. and Henneck, R. and H\'elaine, V. and Hild, N. and Hodge, Z. and Horras, M. and Iaydjiev, P. and Ivanov, S. N. and Kasprzak, M. and Kermaidic, Y. and Kirch, K. and Knecht, A. and Knowles, P. and Koch, H.-C. and Koss, P. A. and Komposch, S. and Kozela, A. and Kraft, A. and Krempel, J. and Ku\ifmmode \acute{z}\else \'{z}\fi{}niak, M. and Lauss, B. and Lefort, T. and Lemi\`ere, Y. and Leredde, A. and Mohanmurthy, P. and Mtchedlishvili, A. and Musgrave, M. and Naviliat-Cuncic, O. and Pais, D. and Piegsa, F. M. and Pierre, E. and Pignol, G. and Plonka-Spehr, C. and Prashanth, P. N. and Qu\'em\'ener, G. and Rawlik, M. and Rebreyend, D. and Rien\"acker, I. and Ries, D. and Roccia, S. and Rogel, G. and Rozpedzik, D. and Schnabel, A. and Schmidt-Wellenburg, P. and Severijns, N. and Shiers, D. and Tavakoli Dinani, R. and Thorne, J. A. and Virot, R. and Voigt, J. and Weis, A. and Wursten, E. and Wyszynski, G. and Zejma, J. and Zenner, J. and Zsigmond, G.},
  journal = {Phys. Rev. Lett.},
  volume = {124},
  issue = {8},
  pages = {081803},
  numpages = {7},
  year = {2020},
  month = {Feb},
  publisher = {American Physical Society},
  doi = {10.1103/PhysRevLett.124.081803},
  url = {https://link.aps.org/doi/10.1103/PhysRevLett.124.081803}
}

@Article{Aggarwal2018,
author={Aggarwal, P. and Bethlem, H. L. and Borschevsky, A. and Denis, M. and Esajas, K. and Haase, P. A. B. and Hao, Y. and Hoekstra, S. and Jungmann, K. and Meijknecht, Th. B. and Mooij, M. C. and Timmermans, R. G. E. and Ubachs, W. and Willmann, L. and Zapara, A.},
title={Measuring the electric dipole moment of the electron in {BaF}},
journal={The European Physical Journal D},
year={2018},
month={Nov},
day={20},
volume={72},
number={11},
pages={197},
issn={1434-6079},
doi={10.1140/epjd/e2018-90192-9},
url={https://doi.org/10.1140/epjd/e2018-90192-9}
}

@article{Aggarwal2021,
    author = {Aggarwal, P. and Bethlem, H. L. and Boeschoten, A. and Borschevsky, A. and Esajas, K. and Hao, Y. and Hoekstra, S. and Jungmann, K. and Marshall, V. R. and Meijknecht, T. B. and Mooij, M. C. and Timmermans, R. G. E. and Touwen, A. and Ubachs, W. and Willmann, L. and Yin, Y. and Zapara, A.},
    title = {A supersonic laser ablation beam source with narrow velocity spreads},
    journal = {Review of Scientific Instruments},
    volume = {92},
    pages = {033202},
    year = {2021},
    number = {03},
    issn = {0034-6748},
    doi = {10.1063/5.0035568},
    url = {https://doi.org/10.1063/5.0035568},
    eprint = {https://pubs.aip.org/aip/rsi/article-pdf/doi/10.1063/5.0035568/13857860/033202\_1\_online.pdf},
}

@Article{Andreev2018,
author={Andreev, V. and
Ang, D. G. and
DeMille, D. and
Doyle, J. M. and
Gabrielse, G. and
Haefner, J. and
Hutzler, N. R. and
Lasner, Z. and
Meisenhelder, C. and
O'Leary, B. R. and
Panda, C. D. and
West, A. D. and
West, E. P. and
Wu, X.
},
title={Improved limit on the electric dipole moment of the electron},
journal={Nature},
year={2018},
day={01},
volume={562},
number={7727},
pages={355-360},
issn={1476-4687},
doi={10.1038/s41586-018-0599-8},
url={https://doi.org/10.1038/s41586-018-0599-8}
}

@article{Boeschoten2024,
  title = {Spin-precession method for sensitive electric dipole moment searches},
  author = {Boeschoten, A. and Marshall, V. R. and Meijknecht, T. B. and Touwen, A. and Bethlem, H. L. and Borschevsky, A. and Hoekstra, S. and van Hofslot, J. W. F. and Jungmann, K. and Mooij, M. C. and Timmermans, R. G. E. and Ubachs, W. and Willmann, L.},
  collaboration = {NL-$e$EDM Collaboration},
  journal = {Phys. Rev. A},
  volume = {110},
  issue = {1},
  pages = {L010801},
  numpages = {5},
  year = {2024},
  month = {Jul},
  publisher = {American Physical Society},
  doi = {10.1103/PhysRevA.110.L010801},
  url = {https://link.aps.org/doi/10.1103/PhysRevA.110.L010801}
}

@phdthesis{BoeschotenThesis,
title = "Precision measurements in diatomic molecules: a route to a permanent electric dipole moment",
author = "Alexander Boeschoten",
year = "2023",
doi = "10.33612/diss.674231809",
language = "English",
publisher = "University of Groningen",
school = "University of Groningen",
}

@article{hao2019,
	author = {Hao, Y. and Pa{\v s}teka, L. F. and Visscher, L. and Aggarwal, P. and Bethlem, H. L. and Boeschoten, A. and Borschevsky, A. and Denis, M. and Esajas, K. and Hoekstra, S. and Jungmann, K. and Marshall, V. R. and Meijknecht, T. B. and Mooij, M. C. and Timmermans, R. G. E. and Touwen, A. and Ubachs, W. and Willmann, L. and Yin, Y. and Zapara, A. and Collaboration), (NL-eEDM},
	doi = {10.1063/1.5098540},
	issn = {0021-9606},
	journal = {J. Chem. Phys.},
	number = {3},
	pages = {034302},
	title = {High accuracy theoretical investigations of {CaF}, {SrF}, and {BaF} and implications for laser-cooling},
	url = {https://doi.org/10.1063/1.5098540},
	volume = {151},
	year = {2019}}

@Manual{WLMWS8-2,
title= {Technical Information Wavelength Meter WS8-2},
OPTauthor= {HighFinesse},
month={Mar},
year = {2021},
url = 
{https://www.highfinesse.com/en/wavelengthmeter/wavelengthmeter-further-information/technical-information-wavelengthmeter-ws8-2.pdf}
}

@phdthesis{MarshallThesis,
title = "Spectroscopy and Systematic Effects: an eEDM experiment using BaF molecules",
author = "Marshall, {Virginia Rose}",
year = "2024",
doi = "10.33612/diss.972290628",
language = "English",
publisher = "University of Groningen",
school = "University of Groningen",
}

@phdthesis{MeijknechtThesis,
title = "Electric and Magnetic Field Control for Electric Dipole Moment Searches",
author = "Thomas Meijknecht",
year = "2023",
doi = "10.33612/diss.822567899",
language = "English",
publisher = "University of Groningen",
school = "University of Groningen",
}

@article{Parker2015,
  title = {First Measurement of the Atomic Electric Dipole Moment of $^{225}\mathrm{Ra}$},
  author = {Parker, R. H. and Dietrich, M. R. and Kalita, M. R. and Lemke, N. D. and Bailey, K. G. and Bishof, M. and Greene, J. P. and Holt, R. J. and Korsch, W. and Lu, Z.-T. and Mueller, P. and O'Connor, T. P. and Singh, J. T.},
  journal = {Phys. Rev. Lett.},
  volume = {114},
  issue = {23},
  pages = {233002},
  numpages = {5},
  year = {2015},
  publisher = {American Physical Society},
  doi = {10.1103/PhysRevLett.114.233002},
  url = {https://link.aps.org/doi/10.1103/PhysRevLett.114.233002}
}

@article{Regan2002,
  title = {New Limit on the Electron Electric Dipole Moment},
  author = {Regan, B. C. and Commins, E. D. and Schmidt, C. J. and DeMille, D.},
  journal = {Phys. Rev. Lett.},
  volume = {88},
  issue = {7},
  pages = {071805},
  numpages = {4},
  year = {2002},
  \\month = {Feb},
  publisher = {American Physical Society},
  doi = {10.1103/PhysRevLett.88.071805},
  url = {https://link.aps.org/doi/10.1103/PhysRevLett.88.071805}
}

@article{Zheng2022,
  title = {Measurement of the Electric Dipole Moment of $^{171}\mathrm{Yb}$ Atoms in an Optical Dipole Trap},
  author = {Zheng, T. A. and Yang, Y. A. and Wang, S.-Z. and Singh, J. T. and Xiong, Z.-X. and Xia, T. and Lu, Z.-T.},
  journal = {Phys. Rev. Lett.},
  volume = {129},
  issue = {8},
  pages = {083001},
  numpages = {6},
  year = {2022},
  publisher = {American Physical Society},
  doi = {10.1103/PhysRevLett.129.083001},
  url = {https://link.aps.org/doi/10.1103/PhysRevLett.129.083001}
}

@article{Dzuba2012,
author = {Dzuba, V. A. and Flambaum, V. V.},
title = {PARITY VIOLATION AND ELECTRIC DIPOLE MOMENTS IN ATOMS AND MOLECULES},
journal = {Int. J. Mod. Phys, E},
volume = {21},
number = {11},
pages = {1230010},
year = {2012},
\\doi = {10.1142/S021830131230010X},
\\URL = {http://www.worldscientific.com/doi/abs/10.1142/S021830131230010X},
\\eprint = {http://www.worldscientific.com/doi/pdf/10.1142/S021830131230010X}
}

@article{Graner2016,
  title = {Reduced Limit on the Permanent Electric Dipole Moment of $^{199}\mathrm{Hg}$},
  author = {Graner, B. and Chen, Y. and Lindahl, E. G. and Heckel, B. R.},
  journal = {Phys. Rev. Lett.},
  volume = {116},
  issue = {16},
  pages = {161601},
  numpages = {5},
  year = {2016},
  month = {Apr},
  publisher = {American Physical Society},
  doi = {10.1103/PhysRevLett.116.161601},
  url = {https://link.aps.org/doi/10.1103/PhysRevLett.116.161601}
}

@article {Jungmann2013,
author = {Jungmann, K.},
title = {Searching for electric dipole moments},
journal = {Ann. d. Physik},
volume = {525},
number = {8-9},
\\url = {http://dx.doi.org/10.1002/andp.201300071},
\\doi = {10.1002/andp.201300071},
pages = {550--564},
year = {2013}
}

@article{Pospelov2005,
title = "Electric dipole moments as probes of new physics",
journal = "Annals of Physics",
volume = "318",
number = "1",
pages = "119 - 169",
year = "2005",
\\doi = "https://doi.org/10.1016/j.aop.2005.04.002",
\\url = "http://www.sciencedirect.com/science/article/pii/S0003491605000539",
author = "M. Pospelov and A. Ritz"
}

@article{Sandars1965,
title = "The electric dipole moment of an atom",
journal = "Phys. Lett.",
volume = "14",
number = "3",
pages = "194 - 196",
year = "1965",
doi = "https:/doi.org/10.1016/0031-9163(65)90583-4",
url = "http://www.sciencedirect.com/science/article/pii/0031916365905834",
author = "P. G. H. Sandars"
}

@article{Sandars1967,
  title = {Measurability of the Proton Electric Dipole Moment},
  author = {Sandars, P. G. H.},
  journal = {Phys. Rev. Lett.},
  volume = {19},
  issue = {24},
  pages = {1396--1398},
  year = {1967},
  publisher = {American Physical Society},
  \\doi = {10.1103/PhysRevLett.19.1396},
  \\url = {https://link.aps.org/doi/10.1103/PhysRevLett.19.1396}
}

@article{Truppe2018,
  title = {A Buffer Gas Beam Source for Short, Intense and Slow Molecular Pulses},
  author = {Truppe, S. and Hambach, M. and Skoff, S. M. and Bulleid, N. E. and Bumby, J. S. and Hendricks, R. J. and Hinds, E. A. and Sauer, B. E. and Tarbutt, M. R.},
  year = {2018},
  journal = {J. Mod. Optics},
  volume = {65},
  number = {5-6},
  pages = {648--656},
  doi = {10.1080/09500340.2017.1384516},
}

@article{Cheng2016,
  title = {Molecular Fountain},
  author = {Cheng, C. and van der Poel, A. P. P. and Jansen, P. and Quintero-P\'erez, M. and Wall, T. E. and Ubachs, W. and Bethlem, H. L.},
  journal = {Phys. Rev. Lett.},
  volume = {117},
  issue = {25},
  pages = {253201},
  year = {2016},
  publisher = {American Physical Society},
  \\doi = {10.1103/PhysRevLett.117.253201},
  \\url = {https://link.aps.org/doi/10.1103/PhysRevLett.117.253201}
}

@article{Haase2021,
  TITLE        = {Systematic study and uncertainty evaluation of {P,T}-odd relativistic enhancement factors in {BaF}},
  AUTHOR       = {P.A.B. Haase and D.J. Doeglas and A. Boeschoten and E. Eliav and M. Ilias and P. Aggarwal and H. L. Bethlem and A. Borschevsky and K. Esajas and Y. Hao and S. Hoekstra and V.R. Marshall and Th. Meijknecht and M. Mooij and K. Steinebach and R.G.E. Timmermans and A. Touwen and W. Ubachs and L. Willmann and Y. Yin},
  year         = {2021},
  journal      = {J. Chem. Phys.},
  Volume  = {155},
  pages = {034309}
  }

@article{Hudson2011,
  title = {Improved measurement of the shape of the electron},
  author = {Hudson, J. J. and Kara, D. M. and Smallman, I. J. and Sauer, B. E. and Tarbutt, M. R. and Hinds, E. A.},
  journal = {Nature},
  volume = {473},
  pages = {493-496},
  year = {2011},
  \\url = {https://www.nature.com/articles/nature10104}
}

@article{Hutzler2012,
author = {Hutzler, N. R. and Lu, H.-I. and Doyle, J. M.},
title = {The Buffer Gas Beam: {A}n Intense, Cold, and Slow Source for Atoms and Molecules},
journal = {Chem. Rev.},
volume = {112},
number = {9},
pages = {4803-4827},
year = {2012},
\\doi = {10.1021/cr200362u},
\\URL = {http://dx.doi.org/10.1021/cr200362u},
\\eprint = {http://dx.doi.org/10.1021/cr200362u}
}

@article{Mooij2024,
doi = {10.1088/1367-2630/ad4207},
url = {https://dx.doi.org/10.1088/1367-2630/ad4207},
year = {2024},
publisher = {IOP Publishing},
volume = {26},
number = {5},
pages = {053009},
author = {Mooij, M. C. and Bethlem, H. L. and Boeschoten, A. and Borschevsky, A. and Esajas, K. and Fikkers, T. H. and Hoekstra, S. and van Hofslot, J. W. F. and Jungmann, K. and Marshall, V. R. and Meijknecht, T. B. and Timmermans, R. G. E. and Touwen, A. and Ubachs, W. and Willmann, L. and Yin, Y. and NL-eEDM collaboration},
title = {Influence of source parameters on the longitudinal phase-space distribution of a pulsed cryogenic beam of barium fluoride molecules},
journal = {New Journal of Physics},
}

@article{Patterson2007,
author = {D. Patterson and J. M. Doyle},
title = {Bright, guided molecular beam with hydrodynamic enhancement},
journal = {J. Chem. Phys.},
volume = {126},
number = {15},
pages = {154307},
year = {2007},
\\doi = {10.1063/1.2717178},
\\URL = {https://doi.org/10.1063/1.2717178},
\\eprint = {https://doi.org/10.1063/1.2717178}
}

@article{Quintero-perez2014,
title = "Preparation of an ultra-cold sample of ammonia molecules for precision measurements",
journal = "J. Mol. Spectr.",
volume = "300",
pages = "112 - 115",
year = "2014",
\\doi = "https://doi.org/10.1016/j.jms.2014.03.018",
\\url = "http://www.sciencedirect.com/science/article/pii/S0022285214000794",
author = "M. Quintero-P\'{e}rez and T. E. Wall and S. Hoekstra and H. L. Bethlem"
}

@article{Roussy2023,
author = {T. S. Roussy  and L. Caldwell  and T. Wright  and W. B. Cairncross  and Y. Shagam  and K. B. Ng  and N. Schlossberger  and S. Y. Park  and A. Wang  and J. Ye  and E. A. Cornell },
title = {An improved bound on the electron’s electric dipole moment},
journal = {Science},
volume = {381},
number = {6653},
pages = {46-50},
year = {2023},
doi = {10.1126/science.adg4084},
URL = {https://www.science.org/doi/abs/10.1126/science.adg4084},
eprint = {https://www.science.org/doi/pdf/10.1126/science.adg4084}}

@article{Sakharov1967,
doi = {10.1070/PU1991v034n05ABEH002497},
url = {https://dx.doi.org/10.1070/PU1991v034n05ABEH002497},
year = {1967},
volume = {5},
pages = {32-35},
author = {A. D. Sakharov},
title = {Violation of {CP} invariance, {C} asymmetry, and baryon asymmetry of the universe},
journal = {Pisma Zh. Eksp. Teor. Fiz.}
}

@article{Shuman2010,
  title = {Laser cooling of a diatomic molecule},
  author = {E. S. Shuman and  J. F. Barry and D. DeMille},
  journal = {Nature},
  volume = {467},
  pages = {820-823},
  year = {2010},
  \\url = {https://www.nature.com/articles/nature09443}
}

@article{Touwen2024,
doi = {10.1088/1367-2630/ad60ee},
url = {https://dx.doi.org/10.1088/1367-2630/ad60ee},
year = {2024},
publisher = {IOP Publishing},
volume = {26},
number = {7},
pages = {073054},
author = {Touwen, A. and van Hofslot, J. W. F. and Qualm, T. and Borchers, R. and Bause, R. and Bethlem, H. L. and Boeschoten, A. and Borschevsky, A. and Fikkers, T. H. and Hoekstra, S. and Jungmann, K. and Marshall, V. R. and Meijknecht, T. B. and Mooij, M. C. and  Timmermans, R. G. E. and Ubachs, W. and Willmann, L. and NL-eEDM collaboration},
title = {Manipulating a beam of barium fluoride molecules using an electrostatic hexapole},
journal = {New Journal of Physics}
}

@article{Vandenberg2014,
title = "Traveling-wave deceleration of {SrF} molecules",
journal = "J. Mol. Spectros.",
volume = "300",
pages = "22 - 25",
year = "2014",
doi = "https://doi.org/10.1016/j.jms.2014.02.004",
url = "http://www.sciencedirect.com/science/article/pii/S0022285214000484",
author = "J. E. van den Berg and S. C. Mathavan and C. Meinema and J. Nauta and T. H. Nijbroek and K. Jungmann and H. L. Bethlem and S. Hoekstra"
}

@article{Engel2013,
	Author = {J. Engel and M. J. Ramsey-Musolf and U. van Kolck},
	Doi = {https://doi.org/10.1016/j.ppnp.2013.03.003},
	Issn = {0146-6410},
	Journal = {Progress in Particle and Nuclear Physics},
	Number = {Supplement C},
	Pages = {21 - 74},
	Title = {Electric dipole moments of nucleons, nuclei, and atoms: The Standard Model and beyond},
	Url = {http://www.sciencedirect.com/science/article/pii/S0146641013000227},
	Volume = {71},
	Year = {2013}
}

@article{Lim2017,
  title = {Laser Cooled {YbF} Molecules for Measuring the Electron's Electric Dipole Moment},
  author = {Lim, J. and Almond, J. R. and Trigatzis, M. A. and Devlin, J. A. and Fitch, N. J. and Sauer, B. E. and Tarbutt, M. R. and Hinds, E. A.},
  journal = {Phys. Rev. Lett.},
  volume = {120},
  issue = {12},
  pages = {123201},
  numpages = {6},
  year = {2018},
  \\month = {Mar},
  publisher = {American Physical Society},
  \\doi = {10.1103/PhysRevLett.120.123201},
  \\url = {https://link.aps.org/doi/10.1103/PhysRevLett.120.123201}
}

@article{Osterwalder2010,
author = {Osterwalder, A. and Meek, S.A. and Hammer, G. and Haak, H. and Meijer, G.},
title = {{Deceleration of neutral molecules in macroscopic traveling traps}},
journal = {Phys. Rev. A},
year = {2010},
volume = {81},
number = {5},
pages = {051401},
month = may
}

@article{Bulleid2012,
author = {Bulleid, N. E. and Hendricks, R. J. and Hinds, E. A. and Meek, S. A. and Meijer, G. and Osterwalder, A. and Tarbutt, M. R.},
title = {{Traveling-wave deceleration of heavy polar molecules in low-field-seeking states}},
journal = {Phys. Rev. A},
year = {2012},
volume = {86},
number = {2},
pages = {021404},
\\month = aug
}

@article{Meerakker2012,
author = {van de Meerakker, S. Y. T. and Bethlem, H. L. and Vanhaecke, N. and Meijer, G.},
title = {{Manipulation and control of molecular beams}},
journal = {Chem. Rev.},
year = {2012},
volume = {112},
number = {9},
pages = {4828--4878}
}

@inproceedings{Boeschoten2023,
  title={Perspectives on electric dipole moments of atoms and molecules},
  author={Boeschoten, A. and Willmann, L.},
  booktitle={EPJ Web of Conferences},
  volume={282},
  pages={01019},
  year={2023},
  organization={EDP Sciences}
}

@article{bause2025,
  title = {Prospects for measuring the electron's electric dipole moment with polyatomic molecules in an optical lattice},
  author = {Bause, R. and Balasubramanian, N. and Fikkers, T. and Prinsen, E. H. and Steinebach, K. and Jadbabaie, A. and Hutzler, N. R. and Aucar, I. A. and Pa\ifmmode \check{s}\else \v{s}\fi{}teka, L. F. and Borschevsky, A. and Hoekstra, S.},
  journal = {Phys. Rev. A},
  volume = {111},
  issue = {6},
  pages = {062815},
  numpages = {15},
  year = {2025},
  publisher = {American Physical Society},
  doi = {10.1103/8ltl-7wsb},
  url = {https://link.aps.org/doi/10.1103/8ltl-7wsb}
}

@article{hofslot2025,
    title={2{D} transverse laser cooling of a hexapole focused beam of cold {BaF} molecules}, 
    author={J. W. F. van Hofslot and I. E. Thompson and A. Touwen and N. Balasubramanian and R. Bause and H. L. Bethlem and A. Borschevsky and T. H. Fikkers and S. Hoekstra and S. A. Jones and J. E. J. Levenga and M. C. Mooij and H. Mulder and B. A. Nijman and E. H. Prinsen and B. J. Schellenberg and L. van Sloten and R. G. E. Timmermans and W. Ubachs and J. de Vries and L. Willmann},
    year={2025},
    journal= {ArXiv},
    pages={2506.19069},
    archivePrefix={arXiv},
    primaryClass={physics.atom-ph},
    url={https://arxiv.org/abs/2506.19069}, 
}

\end{document}